\documentclass[12pt]{article}

\usepackage{placeins}
\usepackage{url}
\usepackage{hyperref}
\usepackage{graphicx}
\usepackage{amsmath}
\usepackage{natbib}
\usepackage[margin=1in]{geometry}
\usepackage{float}
\newcommand{\openpaths}{\textsc{OpenPaths}}

\begin{document}

\title{\openpaths{}: A Supervisor--Specialist Agent System for Personalized, Accessible, and Multi-stop Urban Trip Planning}

\author{Ziyang Xiong, He Zong, Zhiyuan Xue, and Manxi Wu\\
University of California, Berkeley}
\date{}

\maketitle

\begin{abstract}
Urban trip-planning systems are commonly optimized for travel time and cost, but they offer limited support for the heterogeneous needs that real travelers bring, such as personalized preferences, multi-stop itinerary construction, and end-to-end wheelchair accessibility. We present \openpaths{}, a supervisor--specialist multi-agent system that handles all of these tasks within a single architecture. \openpaths{} adopts a deliberate division of labor: LLM agents parse natural-language input, classify request intent, and orchestrate execution, while classical algorithms perform route optimization over curated mobility and accessibility data. This design ensures that the resulting trip honors heterogeneous user preferences and enforces strict accessibility requirements when requested. Beyond per-user planning, \openpaths{} doubles as a measurement instrument for city-scale accessibility analysis: applied to NYC, the system reveals substantial ADA infrastructure gaps and quantifies their effect on job accessibility for wheelchair users. Overall, this study shows how a supervisor--specialist LLM agentic framework can support heterogeneous trip planning and transparent, equitable transportation analysis in real urban environments.

\end{abstract}

\begin{figure}[h]
\centering
    \includegraphics[width=\linewidth]{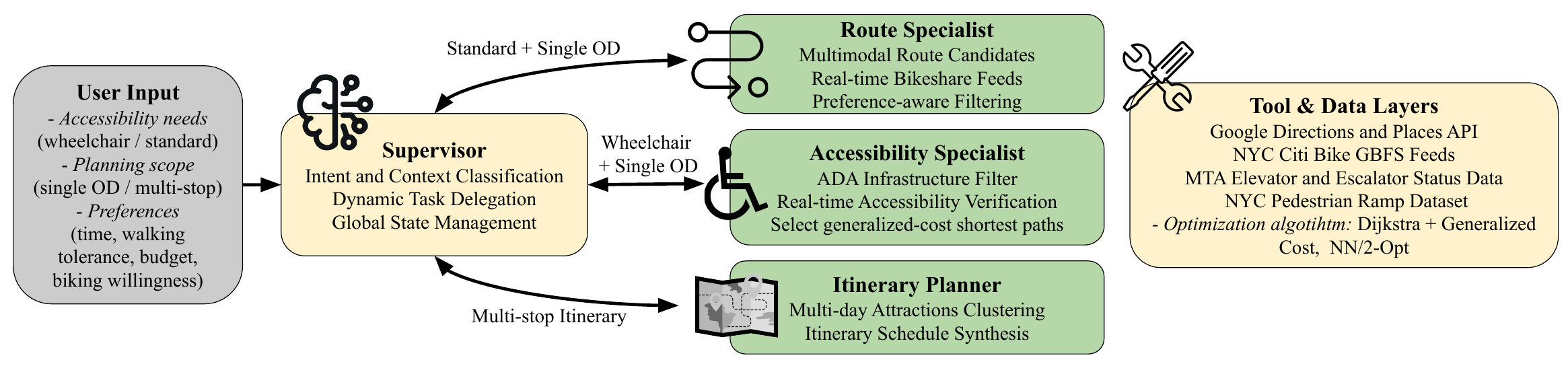}
    \caption{
    \emph{\openpaths{} supervisor--specialist architecture for heterogeneous urban mobility planning.}
The Supervisor Agent acts as the central orchestrator: it parses user input (mobility profile, planning scope, and personalized preferences), classifies query intent, and delegates tasks to three specialists. The Route Specialist handles standard single-OD requests with preference-aware multimodal routing; the Accessibility Specialist handles wheelchair-constrained single-OD requests with ADA-verified routing; and the Itinerary Planner handles single- and multi-day multi-stop trips under either mobility profile. All specialists are grounded in shared urban data and tools, including Google Directions/Places, Citi Bike General Bikeshare Feed Specification (GBFS) feeds, MTA elevator/escalator status data, and pedestrian ramp datasets, with optimization supported by generalized-cost Dijkstra, nearest neighbor, and 2-Opt procedures. Details of this architecture are discussed in Section~\ref{sec:architecture}.
    }
    \label{fig:architecture}
\end{figure}

\section{Introduction}\label{sec:intro}

Digital trip-planning systems such as Google Maps, Apple Maps, and Citymapper have become central to urban mobility, optimizing routes for travel time and cost across multimodal transportation. These systems work well for the most common user request, i.e. the shortest-time directions between two points for an unconstrained traveler, but offer limited support for the heterogeneous planning needs that heterogeneous travelers bring. A single user, across a single trip, may want a route that minimizes walking rather than total time; may want to substitute biking for walking when bikeshare is available nearby; may need to chain six attractions into a coherent two-day itinerary across boroughs without a car; or may need every step of the trip to be wheelchair-accessible end to end. These are different planning problems with different constraints, different data requirements, and different optimization objectives. Most consumer tools handle them in isolation or not at all: preferences such as walking tolerance and biking willingness are not exposed programmatically, multi-stop itinerary planning is largely unsupported, and accessibility filters operate as closed black boxes without auditable infrastructure evidence.

\paragraph{Problem Statement.}
We study urban travel planning across heterogeneous user needs and planning scopes: (i) single-OD routing personalized to each user's stated preferences (walking tolerance, biking willingness, mode and cost weighting), (ii) single-OD routing under strict wheelchair-accessibility constraints, and (iii) multi-stop itinerary construction under either standard or wheelchair-constrained settings. The goal is a single decision-support system that accepts user requests in natural language, handles all of these tasks reliably, returns plans together with human-readable explanations of why each route or itinerary was selected, and exposes its outputs programmatically so the same routing logic can serve a variety of users with heterogeneous preferences and requests.

\paragraph{Our Approach.}
We propose \openpaths{}, a multi-agent planning system organized as a \emph{supervisor--specialist} architecture. A central \emph{Supervisor Agent} parses natural-language constraints, classifies request intent, maintains shared planning state, and delegates to three specialists: a \emph{Route Planning Specialist} that combines Google Directions outputs with real-time bikeshare feeds and reranks routes against the user's stated walking, biking, and time-vs-cost preferences; an \emph{Accessibility Specialist} that ADA-audits transit candidates and computes generalized-cost shortest paths over a curated ADA-filtered station and ramp graph for wheelchair-reliant users; and an \emph{Itinerary Planner} that builds single- and multi-day plans over multiple attractions using Nearest Neighbor + 2-Opt route ordering, with mobility-profile--specific segment synthesis so accessibility constraints carry through to every leg of a multi-stop trip. We adopt a deliberate division of labor: the agent layer handles language understanding, intent-conditional orchestration, and human-readable explanation, while route and itinerary optimization remain classical algorithms grounded in verified accessibility data.

\paragraph{Contributions.}
\begin{enumerate}
    \item \emph{A supervisor--specialist architecture that reliably handles heterogeneous urban-planning tasks.} A single system addresses (i) preference-personalized standard routing, (ii) wheelchair-constrained accessible routing, and (iii) multi-day itinerary construction. The supervisor handles natural-language constraint parsing, intent classification, and delegation; specialists ground LLM reasoning in routing APIs, real-time mobility feeds, and accessibility infrastructure datasets. We report agent-level performance on a labeled benchmark of $M{=}154$ queries spanning all three task types, with intent-classification accuracy of 96.8--98.7\% and end-to-end task success of 95.5\%.

    \item \emph{Auditable accessible routing with end-to-end infrastructure evidence.} Every accessible route returned by \openpaths{} names the specific compliant entrances, ramps, and elevator-equipped stations used at each segment, so accessibility decisions can be inspected, contested, and verified by users, planners, and researchers. Our system treats accessibility as a binding constraint during path construction instead of a post hoc filter. Our design ensures that the system eliminates infeasible segments during generation and extends wheelchair feasibility through every leg of a multi-stop trip.

    \item \emph{City-scale assessment of wheelchair-accessibility in NYC.} Beyond serving individual users, \openpaths{} doubles as a measurement instrument to evaluate the accessibility landscape of a city. Our ADA audit of infrastructure data reveals that only 24.4\% of subway entrances and 30.4\% of pedestrian ramps satisfy compliance criteria. We use \openpaths{} to compute generalized travel costs for accessible and standard routes across all NYC tract pairs and quantify the resulting impact on job accessibility. We find that wheelchair-constrained travelers reach 53.3\% fewer jobs on average within a 45-minute generalized-cost budget, and the loss is spatially concentrated in central Manhattan in absolute terms but in the outer boroughs in relative terms. Together, these results show how \openpaths{} can support both per-user planning and city-scale equity analysis from a single pipeline.
\end{enumerate}

\section{Related Work}\label{sec:related}

\subsection{LLM Agent Frameworks}
Recent work on LLM-based agents demonstrates that language models can decompose complex tasks, invoke external tools, and coordinate multi-step reasoning. Reasoning-and-acting paradigms such as ReAct~\cite{yao2022react} interleave chain-of-thought reasoning with tool calls, while orchestration frameworks such as AutoGen~\cite{wu2024autogen} and LangGraph~\cite{langgraph2024} structure multi-agent workflows around supervisor or graph-based controllers. Travel-domain benchmarks such as TravelPlanner~\cite{xie2024travelplanner} have shown that LLMs alone struggle with the constraint-heavy planning required for realistic itineraries, motivating systems that ground language reasoning in domain-specific tools and verified data. These frameworks provide useful design patterns but are domain-general: they offer the infrastructure for agent coordination without addressing multimodal routing, mobility-aware personalization, or the integration of heterogeneous urban data sources required for transportation planning. \openpaths{} adapts the supervisor--specialist pattern to the transportation setting, with a deliberate division of labor in which LLMs handle natural-language constraint parsing and orchestration while classical algorithms handle optimization over curated mobility and accessibility data.

\subsection{Multimodal and Personalized Routing}
Conventional routing services rank routes by travel time, distance, or generalized travel cost computed over road and transit networks~\cite{google_directions}. Multimodal routing extends this by combining transit, walking, and shared micromobility into single itineraries, and recent work integrates real-time bikeshare and scooter availability to refine walking-heavy segments~\cite{reck2022mode}. Personalization in routing has typically been studied through preference-weighted shortest-path formulations or learned utility models~\cite{letchner2006trip}, but consumer trip planners expose only limited preference controls and do not surface the underlying tradeoffs to users. Multi-stop itinerary problems are classically formulated as variants of the Traveling Salesperson Problem and are commonly addressed with construction-and-improvement heuristics such as Nearest Neighbor and 2-Opt~\cite{lin1973effective}; tourism-recommendation work has extended these formulations with attraction-utility scoring and time-window constraints~\cite{vansteenwegen2011orienteering}. \openpaths{} places personalized multimodal routing, multi-stop itinerary construction, and accessibility-constrained planning under a supervisor-specialist architecture, and grounds optimization in real-time mobility feeds and verified infrastructure data.

\subsection{Accessible Routing and Transit Equity}
Routing for wheelchair-reliant travelers depends on station accessibility, elevator availability, accessible entrances, curb ramps, and transfer burden \cite{mta_data_2024,mta_entrances,nyc_ped_ramps,ada_2010}, and feasible travel requires an unbroken chain of compliant infrastructure. Several research efforts construct accessible routing graphs from open data: AccessMap~\cite{tanweer2017mapping} produces wheelchair-aware sidewalk routes from OpenStreetMap and municipal data, and crowdsourced platforms such as Wheelmap~\cite{wheelmap} catalog the accessibility of individual points of interest. Major consumer navigation services have also begun integrating accessibility filters. Google Maps has integrated a wheelchair-accessible transit option in its mobile application, but these features are not exposed through public APIs, do not produce auditable infrastructure evidence, and do not extend to multi-stop planning. As a result, no existing system supports programmatic accessibility-aware routing combined with trip-level itinerary construction.

Moreover, transportation accessibility research provides the methodological foundation for our city-scale analysis. Accessibility is widely used to evaluate how easily travelers can reach jobs, services, and other opportunities through transport system~\cite{geurs2004_accessibility_review}, and cumulative and cost-based measures have been used to study disparities in opportunity access~\cite{elgeneidy2016_cost_of_equity_tra}. Public-transport studies further document that waiting time, walking time, and transfers impose perceived burdens beyond in-vehicle travel time~\cite{wardman2014_valuing_convenience_itf,schakenbos2016_transfer_valuation}, motivating the generalized-cost formulation we adopt. \openpaths{} extends this literature operationally: rather than evaluating accessibility from precomputed travel-time matrices, it computes wheelchair-feasible generalized-cost paths over a curated ADA-filtered network and exposes the same routing logic to downstream city-scale analysis.

\section{System Architecture}\label{sec:methodology}\label{sec:architecture}

\openpaths{} is organized as a three-layer supervisor--specialist architecture (Figure~\ref{fig:architecture}). The design reflects two observations about accessibility-aware planning. First, user requests arrive as unstructured natural language with implicit constraints--mobility profile, walking tolerance, time and cost budgets, day-level structure for multi-day trips--that must be parsed and routed to the right pipeline. Second, feasibility for wheelchair-reliant travelers depends on combining heterogeneous data sources (directions APIs, real-time mobility feeds, infrastructure inventories). We assign language understanding and orchestration to LLM agents and assign optimization and feasibility checking to deterministic algorithms over curated data, so that route quality and accessibility correctness do not entirely depend on the language model.

\paragraph{Supervisor Layer.}
The Supervisor agent (GPT-4o-mini) converts each request into an executable workflow. It performs (i) \emph{intent and context classification}, identifying whether the request is single-OD routing or multi-stop itinerary planning and detecting accessibility needs; (ii) \emph{dynamic task delegation}, selecting the appropriate specialist(s) and execution order; and (iii) \emph{global state management}, maintaining a shared state object that persists user constraints (e.g., origin/destination, walking tolerance) and intermediate specialist outputs (e.g., candidate route geometries, ADA-compliance verification) across stages.

\paragraph{Specialist Layer.}
Three specialists, each backed by a GPT-4o-family model, perform domain-specific reasoning conditioned on the shared state. The \emph{Route Planning Specialist} (Section~\ref{subsec:standard_routing}) generates multimodal route candidates for standard users, integrating Google Directions outputs with real-time micromobility feeds. The \emph{Accessibility Specialist} (Section~\ref{subsec:accessibility_routing}) computes an ADA-filtered generalized-cost shortest path on the ADA-filtered transit and sidewalk network. The \emph{Itinerary Planner} (Section~\ref{sec:multi_od}) handles multi-stop sequences via Nearest Neighbor + 2-Opt route ordering, with mobility-profile--specific segment synthesis.

\paragraph{Tool and Data Layer.}
Specialists invoke shared tools so LLM reasoning is grounded in real urban data. Navigation is provided by Google Maps Directions/Places; mobility feeds include NYC Citi Bike data \cite{gbfs_station_info}; accessibility infrastructure data includes MTA station equipment and entrance records and the NYC pedestrian ramp inventory. Algorithmic functions implement geocoding, distance computation, generalized-cost Dijkstra search, and itinerary ordering heuristics.


\section{Single-OD Route Planning}\label{sec:single_od}

The Single-OD module produces a feasible route between an origin--destination pair. The Supervisor invokes the Route Planning Specialist for standard users and the Accessibility Specialist for wheelchair-reliant users; both are grounded in shared tools but optimize over different candidate sets.

\subsection{Generalized Travel Cost}
\label{subsec:gc}

Both specialists score paths using a generalized travel cost framework following El-Geneidy et al.~\cite{elgeneidy2016_cost_of_equity_tra}. Each candidate path
$p$ is described by in-vehicle travel time $IVTT(p)$, waiting time $WAIT(p)$, walking and station circulation time $WALK(p)$, number of transfers $NTR(p)$, and fare $FARE(p)$. An example of the generalized cost, expressed in minutes, is
\begin{equation}
\begin{aligned}
GC(p) ={}& 1.0 IVTT(p) + 1.25 WAIT(p) + 1.50 WALK(p)\\
&+ \lambda_{tr} NTR(p) + \frac{FARE(p)}{VOT},
\end{aligned}
\end{equation}
where the 1.25 and 1.50 multipliers reflect the higher perceived burden of waiting and walking time relative to in-vehicle time following transit valuation studies~\cite{elgeneidy2016_cost_of_equity_tra, mta_secondave_feis_appendixD1}, $\lambda_{tr}= 1$~minutes and $VOT = 0.25$~USD/minute following~\cite{mta_secondave_feis_appendixD1}.  These coefficients serve as a default baseline; when the Route Planning Specialist detects user preferences such as walking aversion, transfer aversion, or a different time-versus-cost weighting, it adjusts the corresponding multipliers when re-ranking the available paths.

For each OD pair $(i,j)$ with candidate set $\mathcal{P}_{ij}$, the standard route is $p^{*}_{ij} = \arg\min_{p \in \mathcal{P}_{ij}} GC(p)$. For wheelchair-reliant users, selection is restricted to the accessible subset
\begin{equation}
\mathcal{P}^{acc}_{ij}=\{p\in\mathcal{P}_{ij}\mid Accessible(p)=1\},
\label{eq:acc_set}
\end{equation}
with $C^{wc}_{ij} = \min_{p \in \mathcal{P}^{acc}_{ij}} GC(p)$.

\subsection{Routing for Standard Users}
\label{subsec:standard_routing}

The Route Planning Specialist generates multimodal route options in three stages.

\emph{Candidate retrieval.} The specialist queries the Google Maps Directions API~\cite{google_directions} for the top three multimodal candidates between the origin and destination, ranked by shortest duration. Each candidate is returned as an ordered sequence of walking, transit, and transfer legs with associated duration.

\emph{Preference-conditioned bike augmentation.} The default Google ranking minimizes total duration and can therefore return routes with substantial walking, which some users prefer to avoid. When the user expresses a preference for biking over walking--whether explicitly (e.g., \textit{``I'd rather bike than walk''}) or implicitly via a tight walking tolerance--the specialist scans each candidate for walking legs longer than 10 minutes and queries the NYC Citi Bike GBFS (General Bikeshare Feed Specification) \cite{gbfs_station_info} feed to check real-time station state near each such leg. A bike substitution is generated when there is at least one pickup station with available bikes within 100 meters of the leg's start and at least one drop-off station with available docks within 100 meters of its end. The walking leg is replaced by a walk-to-station / bike-ride / walk-from-station composite, with the new leg duration computed from Citi Bike's reported travel time over the station-to-station segment. Because station status refreshes every 60 seconds, substitutions reflect availability at query time. 

\emph{Preference-aware reranking.} The (possibly augmented) candidate set is then reranked according to the user's stated preferences rather than raw duration. The Supervisor extracts preference parameters from the user's natural-language input--walking tolerance, mode preferences, time-vs-cost weighting--and the specialist scores each candidate against them. Routes whose total walking distance exceeds the user's tolerance are deprioritized even if they are the fastest; routes that better match expressed mode preferences (e.g., bike-augmented variants for bike-friendly users) are promoted. The top-ranked route is returned as the recommendation, with the next-best alternative surfaced for comparison; when no candidate satisfies a hard preference (e.g., the walking tolerance is unmeetable), the violation is explicitly flagged in the response.


\subsection{Accessibility-Aware Routing for Wheelchair-Reliant Users}
\label{subsec:accessibility_routing}


\paragraph{ADA-Compliant Infrastructure Filtering.}
The system first builds a validated infrastructure layer from city datasets. Subway station and entrance records~\cite{mta_entrances} are merged to identify entrances that support wheelchair access. The NYC pedestrian ramp dataset~\cite{nyc_ped_ramps} is filtered using the 2010 ADA Standards~\cite{ada_2010}: ramps must satisfy a maximum running slope of $8.33\%$ ($1{:}12$)~\cite[Section 405.2]{ada_2010} and a minimum clear width of $36$ inches~\cite[Section 405.5]{ada_2010}, and only ramps in acceptable condition are retained. Filtering statistics (Table~\ref{tab:data_filtering}) reveal substantial accessibility gaps in existing infrastructure: only 24.4\% of subway entrances and 30.4\% of pedestrian ramps satisfy ADA criteria.

\begin{table}[htbp]
\centering
\setlength{\tabcolsep}{4pt}
\caption{ADA Filtering Statistics}
\label{tab:data_filtering}
\begin{tabular}{lrrr}
\hline
\emph{Type} & \emph{Original} & \emph{ADA} & \emph{ADA compliant percentage} \\
\hline
Subway Entrances & 1,931 & 472 & 24.4\% \\
Pedestrian Ramps & 219,359 & 66,742 & 30.4\% \\
\emph{Total Nodes} & 221,290 & 67,214 & 30.4\% \\
\hline
\end{tabular}
\end{table}

\paragraph{Generalized-Cost Shortest Path.}
The filtered infrastructure defines an accessible station graph $G = (V, E)$ in which nodes represent ADA-compliant stations and entrances and edges represent feasible station-to-station transitions. The accessibility specialist runs Dijkstra's algorithm on this graph using personalized $GC(p)$ as edge weight. 

\paragraph{Explainability.}
After path selection, the system produces a step-by-step accessible itinerary that names the specific accessible entrances, ramps, and elevator-equipped stations used at each segment, and reports the generalized cost, travel time, and transfer count. This itinerary is included in the Supervisor's final response so users can see why the route was selected and what infrastructure supports it. 

\section{Multi-stop Itinerary Planning}\label{sec:multi_od}

The Itinerary Planner extends the system from point-to-point routing to multi-stop sequences. Given a set of attractions $V = \{v_1, \dots, v_n\}$, optional fixed start $v_{start}$ and end $v_{end}$, and a user-selected optimization criterion (travel time, geographic distance, or monetary cost), the planner produces an ordered sequence $\pi$ minimizing the chosen criterion. We treat ordering as a Traveling Salesperson Problem with optional fixed boundary points.

\paragraph{Visit-Order Optimization.}
Within a single day, we use a two-stage heuristic. \emph{Nearest Neighbor initialization}: starting from $v_{start}$ if specified (otherwise the start that yields the best initial route), the planner repeatedly extends the path to the closest unvisited attraction, producing a feasible initial sequence. \emph{2-Opt local refinement}: the initial route is improved by iterative edge swaps, accepting any swap that improves the objective while leaving fixed boundary points unchanged. The same procedure is applied independently to each day group in multi-day mode.

\paragraph{Multi-Day Day Assignment.}
For multi-day requests, the planner divides attractions into day-level groups $\{V_1, V_2, \dots, V_D\}$ before ordering. When the upstream route output already provides a clear day structure (e.g., user-specified borough-level partitioning), the planner preserves it. Otherwise, attractions are grouped by geographic proximity clustering on pairwise distances $d_{\text{cluster}}(C_i,C_j)=\min_{v_a\in C_i,\,v_b\in C_j} d(v_a,v_b)$, merging the two closest groups iteratively until $D$ groups remain. This keeps day plans spatially coherent and reduces unnecessary cross-borough travel.

\paragraph{Mobility-Aware Synthesis.}
Once the stop order is fixed, segment-level details depend on the user's mobility profile. For each leg $(\pi_i,\pi_{i+1})$, the itinerary for \emph{standard users} is output by the route planning specialist, the itinerary for \emph{wheelchair-reliant users} is provided by the accessibility specialist. The final output provides detailed description of accessible station options and nearby ramps of each transition, so the final travel plan Summary includes both route steps and the accessibility verification behind them.

\section{Evaluation}\label{sec:eval}

We evaluate \openpaths{} along three complementary dimensions: representative case studies that illustrate end-to-end output; an agent-level evaluation of supervisor intent classification and delegation, and a baseline comparison with Google Maps transit API on accessible OD requests.

\subsection{Case Studies}

\subsubsection{Accessible Single-OD Routing}

\begin{figure}[htbp]
    \centering
    \includegraphics[width=\columnwidth]{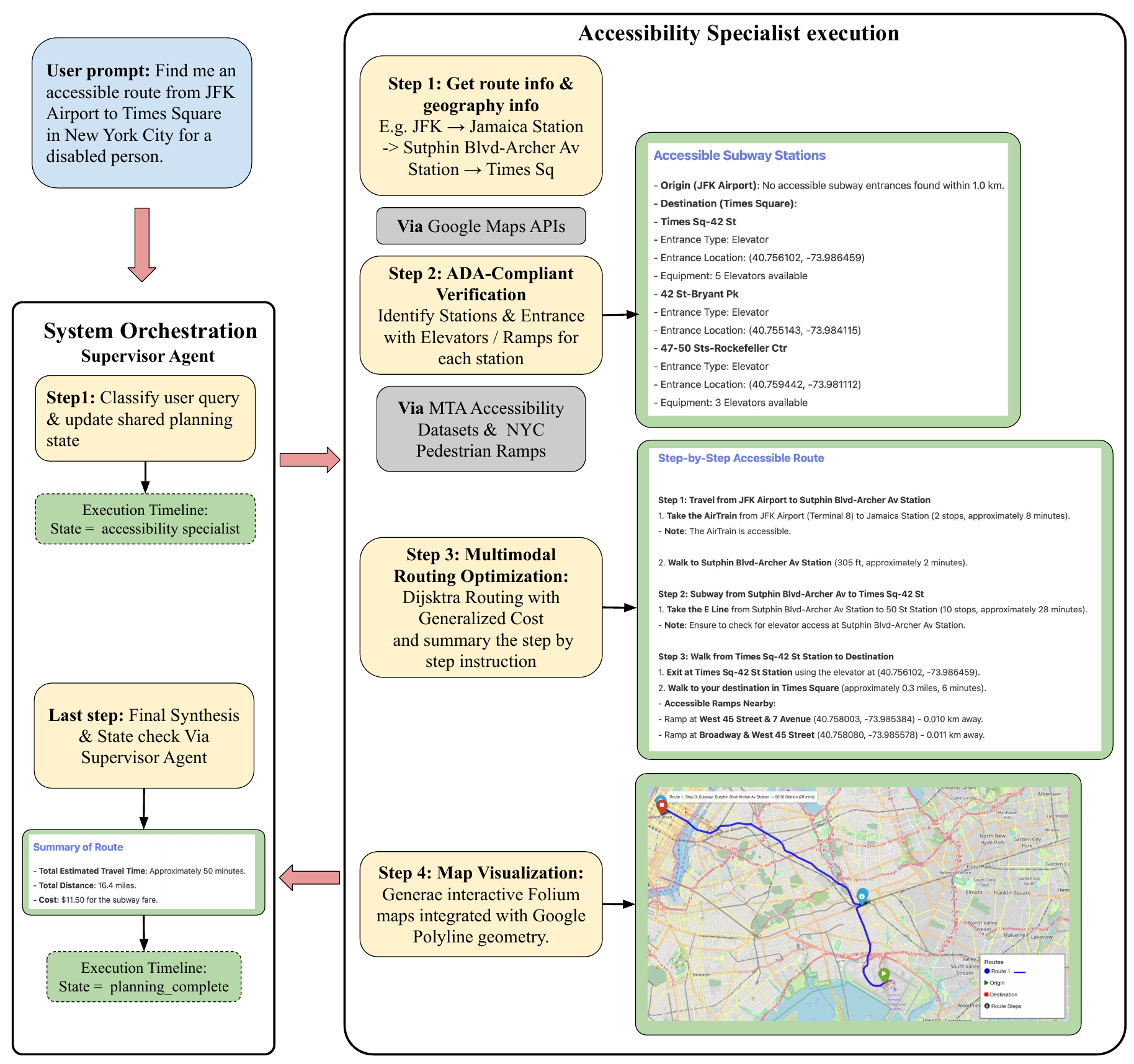}
    \caption{A workflow diagram for wheelchair-accessible single-origin-destination route planning. It shows the supervisor delegating to the accessibility specialist, which retrieves baseline routes, validates them against accessibility datasets, runs generalized-cost path optimization, and returns an audited route for final synthesis.}
    \label{fig:full_workflow}
\end{figure}

Figure~\ref{fig:full_workflow} illustrates the workflow for the query \textit{``Find me an accessible route from JFK Airport to Times Square in New York City for a disabled person.''} The Supervisor parses the request, identifies the accessibility constraint, initializes shared state, and delegates to the Accessibility Specialist. The Specialist applies the generalized-cost Dijkstra procedure of Section~\ref{subsec:accessibility_routing} to select the minimum-burden accessible path. The Supervisor then confirms task completion, audits state consistency, and renders the final route as an interactive map using Folium and Google Polyline geometry.

\subsubsection{Multi-day Itinerary}

\begin{figure}[htbp]
    \centering
    \includegraphics[width=\columnwidth]{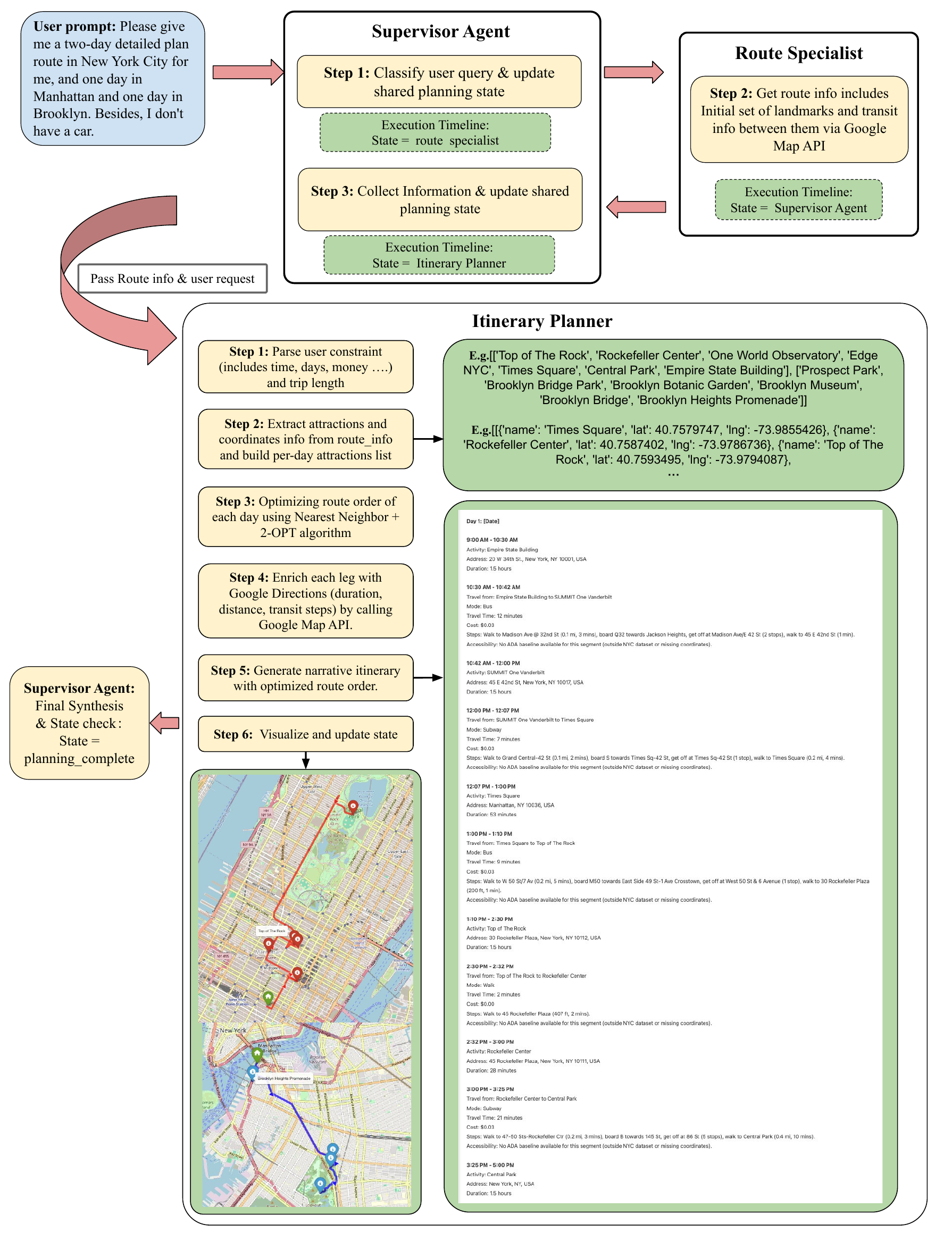}
    \caption{A workflow diagram for multi-day itinerary optimization. It shows the supervisor parsing user constraints, collecting route and landmark information, grouping attractions into day-level clusters, optimizing visit order with nearest neighbor and 2-opt refinement, and synthesizing the final itinerary and map output.}
    \label{fig:multi_day_workflow}
\end{figure}

Figure~\ref{fig:multi_day_workflow} illustrates the workflow for the query \textit{``Please give me a two-day detailed plan route in NYC, one day in Manhattan and one day in Brooklyn. Besides, I don't have a car.''} The Supervisor extracts trip duration, borough preferences, and the transit-only constraint, then invokes the Route Specialist to gather landmark and transit metadata via Google Maps before passing state to the Itinerary Planner. The Planner clusters attractions into a Manhattan day and a Brooklyn day, solves the visit-order problem with nearest neighbor algorithm and 2-Opt algorithm within each day, and enriches the optimized sequence with Google Directions segment details. The Supervisor then synthesizes the final narrative itinerary and accompanying interactive map.

\subsection{Agent-Level Evaluation}
\label{subsec:agent_eval}

To assess whether the Supervisor reliably orchestrates specialist agents, we conduct an agent-level evaluation on a labeled query benchmark.
Rather than relying solely on final outputs, we decompose supervision into interpretable signals, scope and accessibility intent, routing to the appropriate specialists, and end-to-end task completion, so that failures can be localized to planning, delegation, or response generation. 

\paragraph{Dataset construction.}
We first generate candidate travel queries and assign two \emph{gold} labels to each instance:
\emph{request scope} (\texttt{single} vs.\ \texttt{multi}) and
\emph{accessibility need} (\texttt{standard} vs.\ \texttt{wheelchair}).
We then apply a two-stage quality-control pipeline:
(1) automatic screening with \texttt{gpt-4o-mini}, and
(2) manual verification.
This process removes ambiguous or low-quality examples, enforces U.S.-focused scenarios, and re-checks alignment between each query and its \emph{gold}than labels.
The final evaluation set contains \(M=154\) queries.

\paragraph{Execution and logging.}
Each query is executed once through our \openpaths{} Travel Agent system.
For every run, we log the final global state, the invoked agent trace, run status, and final textual output.
We report the following metrics: 
\begin{itemize}
    \item \emph{Intent classification (scope).}
    We measure scope-intent accuracy by checking whether \emph{needs\_itinerary}, as recorded in the final global state, matches gold\_scope.
    \item \emph{Intent classification (accessibility).}
    We measure accessibility-intent accuracy by checking whether \emph{constraints.accessibility}, as recorded in the final global state, matches gold\_accessibility.
    \item \emph{Delegation correctness (self-consistency).}
    We test whether the observed agent trace is consistent with the Supervisor's inferred intent labels (scope and accessibility).
    \item \emph{Delegation correctness (vs gold).}
    We test whether the observed agent trace matches the expected routing pattern implied by \((gold\_scope, gold\_accessibility)\).
    \item \emph{End-to-end task success.}
    A run is evaluated only when \texttt{status=ok} and the output is non-empty; success is then determined by a \texttt{gpt-4o-mini} judge, which provides a concise failure reason for unsuccessful cases.
\end{itemize}

\begin{table}[h]
\centering
\caption{Supervisor performance on the labeled evaluation set (\(M=154\)).}
\label{tab:agent_eval}
\begin{tabular}{lr}
\hline
\emph{Metric} & \emph{Accuracy} \\
\hline
Intent classification (scope) & 96.75\% \\
Intent classification (accessibility) & 98.70\% \\
Delegation correctness (self-consistency) & 97.40\% \\
Delegation correctness (vs gold) & 93.51\% \\
End-to-end task success & 95.45\% \\
\hline
\end{tabular}
\end{table}

The higher self-consistency score (97.40\%) relative to delegation correctness against gold labels (93.51\%) indicates that routing is generally stable once the Supervisor forms an internal intent estimate. The remaining gap is primarily attributable to mismatches between inferred intent and gold intent, rather than to stochastic or inconsistent downstream delegation behavior.

\begin{figure}[htbp]
    \centering
    \includegraphics[width=0.7\columnwidth]{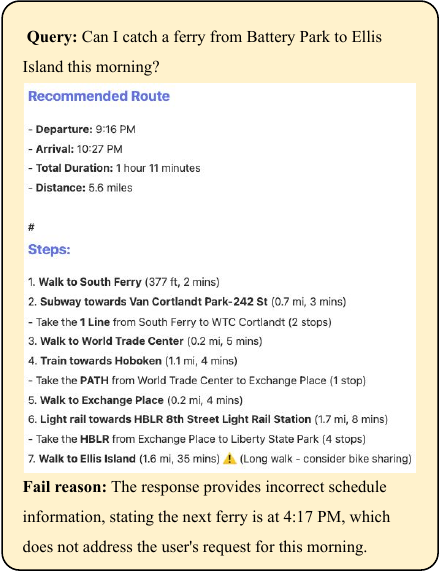}
    \caption{Single-OD time failure: the output follows current-time departures and ignores a user-stated \emph{morning} constraint}
    \label{fig:agent_failure_qualitative}
\end{figure}

\paragraph{Failure case studies.}
By inspecting failed runs in our evaluation logs, we observe several recurring issues, summarized below.
\begin{itemize}
    \item \emph{Accessibility misclassification from ambiguous phrasing.}
    Vague or comfort-oriented wording can bias the Supervisor toward an accessibility-positive label even when gold marks \texttt{standard}.
    For example, \emph{``Help me plan a five-stop culinary tour across New York City that accommodates senior travelers.''} does not request wheelchair access, yet \emph{senior travelers} may be interpreted as mobility-related, incorrectly setting \texttt{constraints} \texttt{.accessibility} and routing through \texttt{accessibility\_specialist}.

    \item \emph{Delegation failure from under-specified multi-day goals.}
    Thematic multi-day requests often bundle only soft preferences; the actionable intent is implicit, so the Supervisor cannot confidently map the utterance to a single next specialist or to a verifiable tool chain.
    Without concrete anchors (dates, seed POIs, explicit OD legs), dispatch rules lack reliable triggers.
    Examples include:

    \emph{``Recommend an outdoor activity schedule in Seattle for a couple of days, considering the rainy weather.''};
    \emph{``Can you suggest a three-day cultural and historical tour in New York City with wheelchair accessibility?''}; 
    \emph{``Organize a three-day exploration in San Francisco, covering both tech museums and outdoor hikes.''}

    When the actionable intent remains implicit, the Supervisor often cannot choose a legitimate next specialist, so dispatch stalls and the graph may terminate immediately after the Supervisor without producing a user-facing plan.

    \item \emph{End-to-end temporal mismatch on single-OD queries.}
    For some single-OD questions, the user states an explicit time window (e.g., \emph{this morning}), but the agent returns results tied to current-time retrieval and ignores the requested time constraint.
    Even when the route structure appears otherwise reasonable, our end-to-end metric still flags these cases as failures because the output violates the user's temporal requirement.
    Figure~\ref{fig:agent_failure_qualitative} shows \emph{``Can I catch a ferry from Battery Park to Ellis Island this morning?''}; the panel reflects current-time departures rather than the user-specified morning window.
\end{itemize}

\subsection{Baseline Comparison with Transit API}
\label{subsec:gmaps_comparison}

We evaluate whether routes returned by a reproducible production routing API can be independently verified as wheelchair accessible. This is not a comparison against the consumer Google Maps wheelchair-accessible mode, which is available in the app but not exposed through the public Directions API. Instead, we use Google transit routing with the documented \texttt{less\_walking} preference as a scriptable baseline and apply the same infrastructure audit to its returned routes. The benchmark starts from 200 NYC OD pairs sampled from census-tract centroids with a fixed random seed and departure time. For \openpaths{}, we evaluate the Accessibility Specialist's routing output under the same OD set.
A returned route is marked ADA verified only if its transit access points and walking endpoints satisfy the filters in Section~\ref{subsec:accessibility_routing}. 

\begin{table}[h]
\centering
\setlength{\tabcolsep}{4pt}
\caption{Auditability check on the 189 OD pairs for which \openpaths{} constructs a verified accessible route.}
\label{tab:gmaps_comparison}
\begin{tabular}{lrrrr}
\hline
\emph{System} & \emph{ADA} & \emph{Walk} & \emph{Time} & \emph{Trans.} \\
 & \emph{(\%)} & \emph{(m)} & \emph{(min)} & \emph{(avg.)} \\
\hline
Google Transit API & 50.3 & 905.8 & 77.3 & 1.5 \\
\openpaths{} & 100.0 & 1083.7 & 96.6 & 1.8 \\
\hline
\end{tabular}
\end{table}

The 200-pair sample is used to define the candidate OD set. The Google Directions baseline returns routes for 197 of these pairs, and 95 of the returned routes pass our audit. \openpaths{} constructs verified accessible route for 189 pairs. We verified that the remaining 11 pairs are not connected via accessible routes, and therefore use these 189 pairs for the auditability check in Table~\ref{tab:gmaps_comparison}. 

Table~\ref{tab:gmaps_comparison} should be read as an auditability baseline rather than a claim that \openpaths{} outperforms the consumer Google Maps accessibility feature. Every \openpaths{} route in this subset passes the accessibility audit. For the same origins and destinations, the Google Directions baseline also returns a route, but only 50.3\% of those routes pass the same audit. The main reason is that a Google itinerary may include at least one transit stop that cannot be matched to an ADA-accessible station or station complex, or a first/last walking endpoint without nearby compliant curb-ramp evidence. The tradeoff is travel burden: \openpaths{} increases average walking distance from 905.8~m to 1083.7~m, travel time from 77.3 to 96.6 minutes, and transfers from 1.5 to 1.8. The result supports a narrower claim: a standard, reproducible routing API can return efficient transit directions, while \openpaths{} returns routes whose accessibility evidence can be explicitly verified end to end.

\section{City-Scale Accessibility Analysis}\label{sec:equity}

The case studies and quantitative comparisons in Section~\ref{sec:eval} establish that \openpaths{} produces feasible accessible routes that match or improve on a production routing service. We now use the system at city scale to characterize the structural accessibility landscape of New York City. Running the accessible routing pipeline across all tract pairs lets us quantify how much job opportunity is lost when accessibility constraints bind, and where those losses concentrate spatially.

\subsection{Accessibility Measure}
\label{sec:cumulative_measure}

For each origin tract $i$, we count reachable opportunities at destination tracts $j$ whose minimum generalized travel cost falls below a threshold $\bar{C}$. Let $C_{ij}^{s}$ denote the minimum generalized cost from $i$ to $j$ under scenario $s \in \{\text{base}, \text{wc}\}$, computed via the Dijkstra procedure of Section~\ref{subsec:accessibility_routing} for the wheelchair-constrained case and over the unrestricted candidate set for the baseline. The cumulative-opportunity accessibility of origin $i$ under threshold $T$ is
\begin{equation}
A_i^{(\bar{C}),s} \;=\; \sum_j O_j \cdot \mathbf{1}\!\left[\, C_{ij}^{s} \le \bar{C} \,\right],
\label{eq:cumulative_access}
\end{equation}
where $O_j$ is the opportunity weight at destination $j$. We use workplace employment from Longitudinal Employer-Household Dynamics Origin-Destination Employment Statistics~\cite{lodes_techdoc_84} as the primary opportunity measure and report results at $\bar{C}=45$ minutes (the main threshold, motivated by NYC's mean commute time of 40.6~minutes~\cite{NYCPlanningPFF}) and $\bar{C}=60$ minutes (a benchmark for robustness). Tract-to-tract costs are approximated by linking each tract centroid to nearby stations and combining first-mile walking access, station-to-station generalized cost, and last-mile walking access, then selecting the minimum-cost connection under each scenario. We adopt the cumulative-opportunity formulation for direct interpretability under fixed travel budgets~\cite{geurs2004_accessibility_review,elgeneidy2016_cost_of_equity_tra}.

\subsection{Jobs-Based Accessibility Results}
\label{sec:jobs_results}

Table~\ref{tab:jobs_accessibility_summary} reports mean tract-level accessibility under both scenarios. At the 45-minute threshold, mean reachable jobs falls from 141{,}060 in the baseline case to 75{,}162 under the wheelchair-constrained case--a 53.3\% reduction averaging roughly 65{,}900 jobs lost per origin tract. The 60-minute threshold raises absolute accessibility in both scenarios, but the relative gap is essentially unchanged (54.5\%, a mean loss of about 178{,}700 jobs). The near-constant relative gap across thresholds is itself informative: extending the travel budget does not help wheelchair users catch up, which suggests the binding constraint is structural. That is, missing accessible infrastructure rather than longer travel distances. This is consistent with the filtering statistics reported in Table~\ref{tab:data_filtering}: only 24.4\% of subway entrances and 30.4\% of pedestrian ramps satisfy ADA criteria, so wheelchair travelers face a sparser network rather than a slower one.


\begin{table}[h]
\centering
\setlength{\tabcolsep}{4pt}
\caption{Mean Tract-Level Jobs Accessibility}
\label{tab:jobs_accessibility_summary}
\begin{tabular}{lrrrr}
\hline
\emph{Thr.} & \emph{Baseline} & \emph{Wheelchair} & \emph{Loss} & \emph{Rel. loss} \\ \hline
45 minutes & 141,059.97 & 75,162.36 & 65,897.61 & 53.25\% \\
60 minutes & 354,649.90 & 175,900.91 & 178,748.98 & 54.54\% \\ \hline
\end{tabular}
\end{table}

\begin{figure}[!t]
    \centering
    \includegraphics[width=\columnwidth]{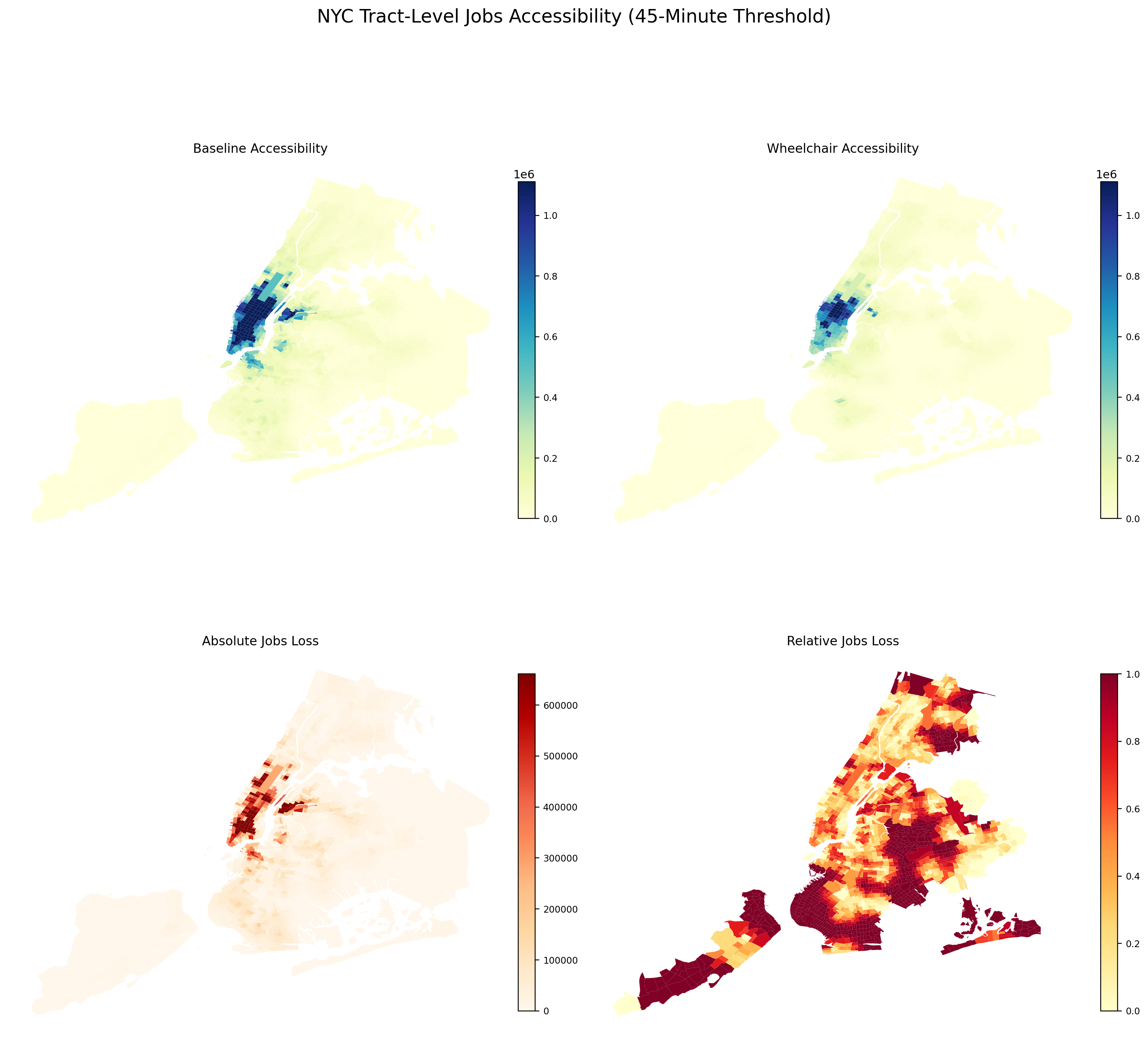}
    \caption{Tract-level jobs-based accessibility maps under baseline and wheelchair-constrained scenarios at the 45-minute threshold. The panels show baseline accessibility, wheelchair-constrained accessibility, absolute jobs loss, and relative jobs loss across New York City census tracts.}
    \label{fig:tract_jobs_heatmap_45}
\end{figure}

Figure~\ref{fig:tract_jobs_heatmap_45} shows that this gap is spatially uneven, and the absolute and relative loss panels tell complementary stories. Absolute losses concentrate in central Manhattan and adjacent corridors with strong transit centrality--the tracts that lose the most jobs are precisely those best-served by the standard network, where the gulf between potential and accessible reach is the widest. The relative-loss panel inverts this geography: while Manhattan exhibits moderate-to-high relative losses, the outer boroughs--large portions of the Bronx, eastern Queens, southern Brooklyn, and nearly all of Staten Island--register near-total relative loss in the 80--100\% range, indicating that whatever job access these tracts had under the standard network is almost entirely erased once accessibility constraints bind. Representative high-loss origin tracts are reported in Table~\ref{tab:top_loss_tracts}. Taken together, the two panels identify two distinct populations of concern: wheelchair residents in central Manhattan, where strong nominal access masks a large effective gap, and residents of outer-borough tracts, where the remaining accessible network is fragmentary to the point of near-total exclusion. The pattern also identifies asymmetric retrofit leverage: closing a small number of accessibility gaps in the densest job corridors could recover a disproportionate share of the absolute loss, while improving baseline access in the most-isolated outer-borough tracts addresses the most severe relative exclusion. \openpaths{} surfaces both lenses simultaneously, showing how a single tool can support both efficiency-oriented and equity-oriented infrastructure planning.


\begin{table}[h]
\centering
\caption{Representative High-Loss Origin Tracts}
\label{tab:top_loss_tracts}
\begin{tabular}{lp{0.42\columnwidth}r}
\hline
\emph{Threshold} & \emph{Neighborhood / Tract} & \emph{Jobs Loss} \\ \hline
45 minutes & Greenwich Village (CT 57) & 1,433,917 \\
45 minutes & Midtown South--Flatiron--Union Sq. (CT 52) & 1,382,120 \\
45 minutes & Greenwich Village (CT 61) & 1,354,486 \\
60 minutes & Lower East Side (CT 14.02) & 1,751,461 \\
60 minutes & Lower East Side (CT 14.01) & 1,734,761 \\
60 minutes & Chinatown--Two Bridges (CT 6) & 1,649,840 \\ \hline
\end{tabular}
\end{table}

\FloatBarrier

\section{Conclusion}\label{sec:conclusion}
We presented \openpaths{}, a supervisor--specialist multi-agent system for urban mobility planning that handles preference-personalized routing, wheelchair-constrained accessible routing, and multi-stop itinerary construction within a single architecture. \openpaths{} adopts a deliberate division of labor: LLM agents handle natural-language constraint parsing, intent classification, orchestration, and human-readable explanation, while classical algorithms handle optimization over curated mobility and accessibility data. Each route or itinerary is returned with the evidence behind it, including the named infrastructure used for accessible trips, so recommendations can be inspected and verified rather simply accepting the output route.

Beyond per-user planning, the city-scale analysis in Section~\ref{sec:equity} demonstrates that \openpaths{} doubles as a measurement instrument for urban accessibility: applying the system to NYC reveals that only 24.4\% of subway entrances and 30.4\% of pedestrian ramps satisfy ADA criteria, and that wheelchair-constrained travelers reach 53.3\% fewer jobs on average within a 45-minute cost budget, with losses concentrated in central Manhattan in absolute terms and in the outer boroughs in relative terms. These structural disparities are difficult to observe from standard routing outputs alone, and they identify where infrastructure retrofits would have the highest leverage.

Future work will extend the framework along three directions. First, we plan to generalize \openpaths{} beyond New York City by ingesting standardized mobility and accessibility datasets from additional metropolitan areas. Second, we aim to incorporate participatory and real-time feedback, letting users report infrastructure disruptions such as elevator outages and blocked ramps so the system maintains a dynamic urban accessibility model. Third, we will broaden the planning scope to include pricing, activity scheduling, and hospitality coordination. Overall, \openpaths{} suggests that personalized, accessible, and multi-stop urban trip planning can be treated as a unified urban computing problem in which the same agent-orchestrated pipeline supports both individual decision-making and city-scale equity analysis grounded in heterogeneous urban data.

\bibliographystyle{ACM-Reference-Format}
\bibliography{references}

\end{document}